\documentclass[runningheads]{llncs}
\usepackage[]{todonotes}
\usepackage{listings}
\usepackage{subcaption}
\usepackage{tikz}
\usepackage{pgfplots}
\usepackage[final]{changes}
\usetikzlibrary{patterns}
\begin{document}
\hyphenation{block-chain block-chains Bul-let-proofs pro-vid-er pro-vid-ers off-loading}
\title{Blockchain-based Result Verification for Computation Offloading}
\titlerunning{Blockchain-based Result Verification for Computation Offloading}

\author{Benjamin Körbel \and Marten Sigwart \and Philip Frauenthaler \and Michael Sober \and Stefan Schulte\orcidID{0000-0001-6828-9945}}
\authorrunning{B. Körbel et al.}
%
\institute{Christian Doppler Laboratory for Blockchain Technologies for the Internet of Things \\ TU Hamburg, Germany; TU Wien, Austria\\
\email{\{michael.sober|stefan.schulte\}@tuhh.de}\\
\url{https://www.cdl-bot.at}}

\maketitle              
\noindent
\fbox{\parbox{\linewidth}{
		NOTICE: This is the authors' version of the accepted manuscript at the 19th International Conference on Service Oriented Computing. Please cite as: \textbf{Benjamin Körbel, Marten Sigwart, Philip Frauenthaler, Michael Sober, Stefan Schulte: Blockchain-based Result Verification for Computation Offloading. 19th International Conference on Service Oriented Computing (ICSOC), 2021}. 
}}

\begin{abstract}
Offloading of computation, e.g., to the cloud, is today a major task in distributed systems. Usually, consumers which apply offloading have to trust that a particular functionality offered by a service provider is delivering correct results. While redundancy (i.e., offloading a task to more than one service provider) or (partial) reprocessing help to identify correct results, they also lead to significantly higher cost.

Hence, within this paper, we present an approach to verify the results of offchain computations via the blockchain. For this, we apply zero-knowledge proofs to provide evidence that results are correct. Using our approach, it is possible to establish trust between a service consumer and arbitrary service providers. We evaluate our approach using a very well-known example task, i.e., the Traveling Salesman Problem.
\end{abstract}
\section{Introduction}
\label{sec:intro}
Offloading of computational tasks has gained a lot of research attention in recent years~\cite{DBLP:journals/sigmobile/Satyanarayanan14}. 
The basic idea of offloading is that a client device outsources resource-intensive computational tasks to providers, often in exchange for a fee~\cite{DBLP:conf/hicss/MarstonLBG11}. Hence, when offloading tasks, two parties are involved. Task issuers (i.e., \emph{service consumers}) potentially have limited computational capabilities and therefore are interested in outsourcing particular tasks. Conversely, task processors (i.e., \emph{service providers}) may have idle computational resources and offer their CPU-cycles and further computational resources to process these tasks. 
Typical examples are the offloading of data processing tasks from lightweight Internet of Things (IoT) or mobile devices in order to decrease processing time or to save energy~\cite{wu20}. For instance, machine learning, (combinatorial) optimization tasks, or the application of heuristics (e.g., genetic algorithms) to solve a complex problem require often resources not available to a potential service consumer. Apart from overcoming limited computational resources, scalability and fault tolerance are also major reasons why offloading is applied~\cite{kosta12}.

Offloading can be done to resources following the Infrastructure-as-a-Service (IaaS), Platform-as-a-Service~(PaaS), or Software-as-a-Service~(SaaS) models, depending on the needs of the service consumer. Traditionally, computation offloading leverages the cloud, e.g.,~\cite{khan14}, but more recently, offloading to resources at the edge of the network has also been widely discussed, e.g.,~\cite{mach17}. 

Regardless of the technological setting, offloading requires a client to trust the service provider to deliver correct results. This is a major market entry barrier, since service consumers naturally trust well-known service providers more than new market participants. \deleted{Very often, this also leads to vendor lock-in, since service consumers become technology-bound to a particular provider~\cite{satzger13}.}

In order to avoid reliance on a particular provider, the usage of blockchain technologies for task offloading has previously been discussed~\cite{DBLP:journals/csm/UriarteN18}. In such approaches (e.g., Golem or iExec---see Section~\ref{sec:related}), the blockchain is a service broker, which brings together consumers and providers, and often delivers further functionalities, e.g., automated settlement after the offloading task has been carried out. Also, the offloading results are delivered through the blockchain.

However, to the best of our knowledge, none of the existing approaches performs a check of the correctness of the delivered results. Ideally, before service consumers pay the service providers for their work, they have an assurance that the returned results can be fully trusted. Previous studies are aware of this issue and discuss solutions based on, e.g., redundant computing, reprocessing fractions of a task locally, or reputation-based systems in order to ensure correct results~\cite{DBLP:journals/tmc/ChatzopoulosAKH18,iexecWhitepaper,golemArchitecture}. While these approaches may reduce the risk of receiving wrong results, they cannot proof that a result is correct~\cite{DBLP:journals/csm/UriarteN18}.
In other, non-blockchain solutions, the user needs to trust a third party which provides functionalities ensuring trust in the offloading results, e.g.,~\cite{santos09}.

\added{Furthermore, it should be noted that many offloading tasks do not deliver a deterministic result. For instance, if offloading machine learning tasks or heuristics to solve NP-complete problems, the computation results can differ. This further complicates checking the correctness of a result, since redundant computing or partial reprocessing may lead to different results.}

Within the work at hand, we address this issue by conceptualizing, implementing, and evaluating a blockchain-based offloading approach that can prove the proper execution of a particular computation task. By using a blockchain\deleted{as underlying infrastructure}, we dissolve the dependency on a trusted third party. Using a public blockchain also helps to achieve transparency, since information about the off-chaining procedure is publicly available. To prove the correctness of computational results, we apply zero-knowledge proofs (ZKPs). We evaluate our approach using the well-known Traveling Salesman Problem (TSP), showing in which use case areas the proposed solution is beneficial if compared to other alternative approaches, and assessing the cost and time overhead of our approach.

\added{In brief, we provide the following contributions in this paper:}
\begin{itemize}
	\item \added{We assess approaches to ensure trust in results provided by service providers.}
	\item \added{We discuss the utilization of ZKPs and blockchain technologies in order to verify the results for offloaded tasks.}
	\item \added{We design and implement a blockchain-based solution for computation offloading with result verification.}
	\item \added{We apply the TSP as a running example and in order to evaluate the overhead resulting from the presented approach.}
\end{itemize}
The remainder of this paper is organized as follows: In Section~\ref{sec:related}, we discuss the related work. In Section~\ref{sec:offchain}, we assess different approaches to verify the results of offloaded computation tasks. Based on this, we present our design and implementation in Section~\ref{sec:design}. Section~\ref{sec:evaluation} shows the results of the evaluation of the presented work, and Section~\ref{sec:conclusion} concludes this paper.


\section{Related Work}
\label{sec:related}
To the best of our knowledge, the field of blockchain-based, verifiable task offloading is still a novel research area, and not too many approaches have been presented so far. 

Golem~\cite{golemArchitecture}, iExec~\cite{iexecWhitepaper}, and SONM~\cite{sonmWhitepaper} are three commercial solutions, aiming at decentralizing offloading to the cloud~\cite{DBLP:journals/csm/UriarteN18}. Their respective primary goal is to provide solutions to decrease market entry barriers, by allowing arbitrary providers to offer computational resources on a blockchain, and arbitrary consumers to use these resources. Notably, the intended providers of cloud resources are not large-scale data centers, but could be anyone with idle computational resources. 
Golem, iExec and SONM aim at providing marketplace and broker functionalities, and apply a pay-per-use model, i.e., the consumer has to pay for using computational infrastructure or for processing a particular task. \deleted{All three solutions are based on Ethereum.}
Notably, in contrast to the work at hand, which focuses on a SaaS model, these solutions aim at providing computing power in general, i.e., on the IaaS level. 

With regard to result verification, Golem supports redundant computation, but also allows to recompute fractions of an offloaded task locally (i.e., at the service consumer's side), and to subsequently compare the results. 
Also, Golem implements a reputation mechanism, which is based on consumer (e.g., late payments) and provider behavior (e.g., not delivering results in time), respectively~\cite{golemArchitecture}. iExec applies a similar approach, where the service consumer can define the needed reliability of the results. If this value is high, a higher degree of redundancy is applied when computing the offloading tasks, and more reliable providers, i.e., with a high reputation, are selected. Notably, iExec also allows to support Software Guard Extension (SGX), which is a kind of enclave-based off-chain computations (see Section~\ref{sec:offchain})~\cite{iexecWhitepaper}. So far, SONM does not implement a verification mechanism, but names reputation management as a major enabler to provide reliable computation results~\cite{sonmWhitepaper}.

None of the so-far discussed approaches provides a proof that the results of a computation are correct. Instead, redundant computations, recomputing fractions of tasks locally, and reputation-based methods only \emph{decrease} the risk that the results are not correct. Especially redundant computations also increase the cost by quite some degree, since all involved service providers charge a fee for the computations. Reputation systems can be helpful, but provide market entry barriers since new service providers need to build a reputation. Also, it remains unclear how these solutions handle results which are not deterministic, e.g., for machine learning or heuristic tasks. 


FlopCoin~\cite{DBLP:journals/tmc/ChatzopoulosAKH18} is a blockchain-based offloading framework with a decentralized incentive and reputation scheme. Among other metrics, the reputation of participants is used as input for the offloading decision, i.e., to which provider of computational resources a particular task is offloaded. \deleted{FlopCoin primarily aims at offloading from mobile devices.} 
EdgeChain~\cite{DBLP:journals/iotj/PanWHALZ19} uses a blockchain and smart contracts to link computational resources at the edge and IoT devices which need to offload tasks. 
The blockchain is used to monitor the offloading procedure and to conduct payments. A mechanism to detect malicious nodes based on past behavior is also introduced. Qiu et al.~\cite{DBLP:journals/tvt/QiuLCHZ19} discuss a similar approach, but apply deep reinforcement learning to find an assignment of tasks and available edge resources. Very recently, another approach for offloading to the fog has been presented by Wu et al.~\cite{wu21}. \deleted{Here, the authors make use of the blockchain in order to ensure data integrity during the offloading process.} The focus of this work is also on the actual decision making, i.e., where to place which offloaded task. In contrast, we aim primarily on proving that computed results are valid. 

\deleted{Tang et al.~\cite{DBLP:conf/ispa/TangZRD18} present offloading for smart vehicles, also applying a blockchain for coordinating, logging, and monitoring the offloading process. Interestingly, the authors additionally focus on security aspects, using the blockchain to verify the authenticity of the provided computational resources, but do not prove the correctness of results.} 

To the best of our knowledge, none of the discussed research papers directly address offloading result verification. Hence, the work at hand complements existing work, and could be used within existing solutions in order to proof that an offloaded computation provides valid results.
\section{Result Verification for Offloaded Tasks}
\label{sec:offchain}
As discussed above, it is the goal of the work at hand to provide mechanisms that can verify results of offloaded computational tasks. In general, we focus on the SaaS model, but in fact, result verification could also be done for user-deployed services using the IaaS or PaaS model.  

To achieve result verification, different schemes could be applied: \textit{Verifiable off-chain computation} entails the provisioning of cryptographic proofs that witness correct processing. After a computation is performed, a cryptographic proof is generated and published together with the result on a blockchain by the processor (here: the service provider). Subsequently, the validity of the computation can be verified on-chain using a smart contract~\cite{DBLP:conf/middleware/EberhardtH18}.

Verifiable off-chain computation can be realized using ZKPs~\cite{DBLP:conf/stoc/GoldreichMW87}. The basic idea behind ZKPs is to convince someone that a statement is true without revealing any underlying information needed to proof that the statement is true. This allows to hide the input data for a proof and therefore supports data privacy. \added{Importantly, in the scenario at hand, this facilitates the verification that the results delivered by a service provider are correct, without the need that the provider reveals its applied service or algorithm. Hence, the computation performed by the service provider remains a blackbox from the perspective of result verification. This is even the case for non-deterministic computations, e.g., if a heuristic is applied. As long as it is possible to define rules which describe if an offloading result is valid, ZKPs can be applied successfully.}

ZKPs can be realized in the form of Zero-Knowledge Non-Interactive Succinct Arguments of Knowledge (zk-SNARKs), Zero-Knowledge Scalable Transparent Arguments of Knowledge (zk-STARKs), and Bulletproofs \cite{kosba20}. 

zk-SNARKs are non-interactive and provide relatively cheap verification by their succinctness. 
Before generating a proof and performing the verification step, a one-time setup must be carried out by a trusted party. 
Unlike zk-SNARKs, zk-STARKs and Bulletproofs do not require a trusted one-time setup. In zk-SNARKs and Bulletproofs, computations are abstracted with arithmetic circuits, while zk-STARKs leverage higher degree polynomials. \deleted{As a consequence, no tools are available to specify programs for zk-STARKs~\cite{DBLP:conf/middleware/EberhardtH18}, while according tools for zk-SNARKs have been presented~\cite{DBLP:conf/ithings/EberhardtT18,DBLP:conf/sp/KosbaPS18}.} Both zk-STARKs and Bulletproofs feature growing proof-size and on-chain verification, while zk-SNARKs are independent of the task complexity and provide compact proves~\cite{DBLP:conf/middleware/EberhardtH18}. Due to the succinctness of zk-SNARKs, very short proofs (i.e., in the range of bytes) can be provided, which is very beneficial when blockchain technology is involved. Therefore, we decided to apply zk-SNARKs for result verification. 

We have also investigated other result verification schemes: For instance, \emph{Secure Multiparty Computation} (SMPC) protocols enable the construction of privacy-preserving off-chain computation schemes, but are accompanied by high overhead~\cite{DBLP:conf/middleware/EberhardtH18}. \emph{Enclave-based off-chain computation} relies on Trusted Execution Environments (TEEs) which enable code execution while preserving confidentiality and integrity. The enclave-based scheme allows universal computations but has potential security issues~\cite{intel2019Plundervolt,DBLP:journals/corr/SchwarzWGMM17}. \emph{Incentive-driven off-chain computing} rewards nodes which are doing verification work to check if a computation is correct. One implementation of this scheme is TrueBit~\cite{DBLP:journals/corr/abs-1908-04756}. A challenge when using the described scheme is to keep nodes motivated for performing verifications continuously. Also, the throughput of completed computation tasks and the general service can be hindered by malicious verifiers by marking each computation result as faulty~\cite{DBLP:conf/middleware/EberhardtH18}.

The selection of zk-SNARKs allows us to make use of the ZoKrates toolbox~\cite{DBLP:conf/ithings/EberhardtT18}, which supports the entire process of specifying, integrating and deploying ZKPs on Ethereum-based blockchains. The toolbox consists of a Domain-specific Language (DSL), a compiler and generators for proofs as well as smart contracts for verification. 
In brief, ZoKrates can be used to execute a computational task off-chain. Afterwards, the result of the computation (here: of an offloading task) and the corresponding proof are written back on a blockchain. The proof that attests correct (or incorrect) computation can then be verified on-chain. Therefore, the computational effort on a blockchain is reduced, while privacy can be preserved due to the usage of ZKPs.

\section{Design and Implementation}
\label{sec:design}
\subsection{Overview}
\label{sub:designoverview}
After having selected zk-SNARKs as the underlying approach to provide result verification for computation offloading, we are now able to design a solution. 

As discussed before, we make use of a blockchain-based approach to offload tasks to service providers. While this has been proposed before (see Section~\ref{sec:related}), there is lack of solutions which allow to verify the results delivered by the service providers. Due to space constraints, we focus on this particular functionality in the work at hand. However, we have in fact designed and implemented a framework which covers the necessary functionality stack, i.e., acts as a blockchain-based broker for service consumers (offloaders) and service providers, and implements an incentive structure, so that fees can be charged and are automatically paid if a result has been verified. Notably, while the implemented solution can be used by traditional cloud providers to offer their resources and services, it could also be used in fog and edge settings, or by private persons who want to offer spare computational resources.  

In the case of non-deterministic results, e.g., since a heuristic is applied by a service provider (see Section~\ref{sec:intro}), our framework allows to obtain results from different providers and to compare the result quality. The integration of methods to assess the result quality is part of our future work (see Section~\ref{sec:conclusion}).

\begin{figure}[tb] 
	\centering
	\includegraphics[width=1.0\textwidth]{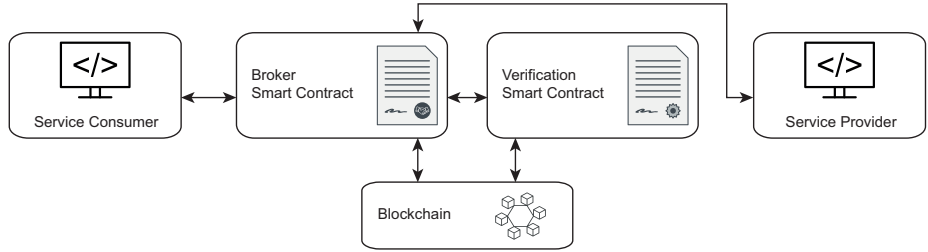}
	\caption{Blockchain-based computation offloading with result verification}
	\label{fig:overview}
\end{figure}

Figure~\ref{fig:overview} shows the components of the software solution. As it can be seen, the system consists of the service consumer (i.e., the client software for the task offloader), the service provider (i.e., the according client software), the broker smart contract and the result verification smart contract. 
In the following subsections, we discuss the core components with a focus on the verification functionalities.

\subsection{\replaced{Blockchain-based Brokering and Result Verification}{Blockchain as Infrastructure}}
\label{sub:brokering}
The blockchain serves two major purposes in our scenario: First, it acts as a broker during the offloading process, i.e., facilitates the cooperation between a service consumer and a service provider. This includes provisioning of results to the service consumer and payment to the service provider
. Second, the blockchain delivers the result verification (see Section~\ref{sub:resultverification}). Following the approach presented in this work, no preexisting relationship and no position of trust between a service consumer and potential service providers need to exist. 

Brokering functionalities and result verification are implemented using smart contracts. 
Within the work at hand, we use Ethereum for this, since it provides a broad acceptance in the research community as well as industry. It should be noted that the presented approach is per se protocol-agnostic, and could also be implemented using a permissioned blockchain like Hyperledger Fabric. However, we opted to use a public blockchain in the work at hand.
\subsection{Broker Smart Contract}
\label{sub:brokersmartcontract}
\begin{figure}[h] 
	\centering
	\includegraphics[width=0.85\textwidth]{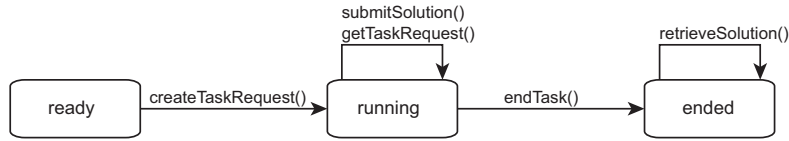}
	\caption{Simplified state diagram of the broker smart contract}
	\label{fig:auctionSmartContractStates}
\end{figure}

Figure~\ref{fig:auctionSmartContractStates} provides an overview of the states the broker smart contract passes through. As it can be seen in the state diagram, the contract may be in the states \emph{ready}, \emph{running}, and \emph{ended}. For the transition between the states, particular functions must be called. Before any interaction is possible, i.e., any function is callable, the smart contract needs to be deployed on the blockchain. After the deployment, it is in state \emph{ready}. At this point, a service consumer can create a new request for offloading, i.e., a new offloading task, using \emph{createTaskRequest()}.

The necessary inputs for \emph{createTaskRequest()} are a stake, which is used as a deposit for the later payment to the service provider, information about the offloaded task, and a boolean value if the result should be verified or not. Once this has been done, the state changes to \emph{running}.

At this point of time, a potential service provider can retrieve all the necessary information to process the task by calling \emph{getTaskRequest()}. Notably, the selection of the service provider could follow different patterns, e.g., based on reputation and/or load balancing as proposed in the related work (see Section~\ref{sec:related}), by applying a reverse auction so that potential service providers compete for the requests, or other allocation techniques. 

Since this is not in the focus of the work at hand, we implement a simplified approach, i.e., 
once a service provider has computed a result, it can be published using \emph{submitSolution()}. 
When calling this function, the service provider needs to deliver the actual solution to the request. At this point, network participants including the service consumer can see the submitted solution due to the public nature of the blockchain. To circumvent that a service consumer reads the result and does not pay the provider, the stake deposited by the consumer is used. 

Any network participant (here: the service consumer or the service providers) may close the task by using \emph{endTask()}, which also means that the state changes to \emph{ended} and that the payment to the service provider is triggered. 
Notably, this is only possible once a minimum duration has passed, which is also defined by the service consumer. 
When the task is ended, the service consumer can collect the solution by calling the function \emph{retrieveSolution()}. As written before, the consumer could also read the result simply from the blockchain. We explicitly foresee \emph{retrieveSolution()} as the possibility to implement a more sophisticated function here, e.g., to encrypt and decrypt the result in order to not release the solution publicly on the blockchain, or to provide the solution via a blockchain-external data storage like the InterPlanetary File System (IFPS)~\cite{KSS20}, in order to save gas cost. 

\subsection{Result Verification}
\label{sub:resultverification}

\begin{figure}[t] 
	\centering
	\includegraphics[width=\textwidth]{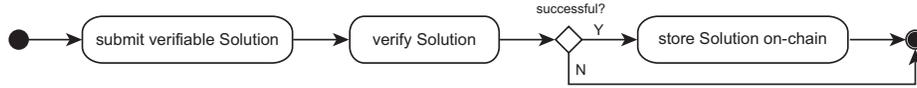}
	\caption{High level process of the result verification}
	\label{fig:flowResultVerification}
\end{figure}

Figure~\ref{fig:flowResultVerification} gives a high level overview of result verification within our approach. As it can be seen, solutions of verifiable tasks are written to the blockchain only if the result verification is successful. Otherwise, the submission is discarded. 
The activity~\textit{verify Solution} is part of the result verification while all other activities in Figure~\ref{fig:flowResultVerification} belong to the broker smart contract discussed in Section~\ref{sub:brokersmartcontract}. The verification is performed within a separate smart contract generated by ZoKrates. Therefore, the broker smart contract has to trigger the verification function \textit{verifyTx()} in the verification smart contract. An example implementation of \emph{verifyTx()} is discussed in Section~\ref{sub:example}.

\begin{figure}[t] 
	\centering
	\includegraphics[width=\textwidth]{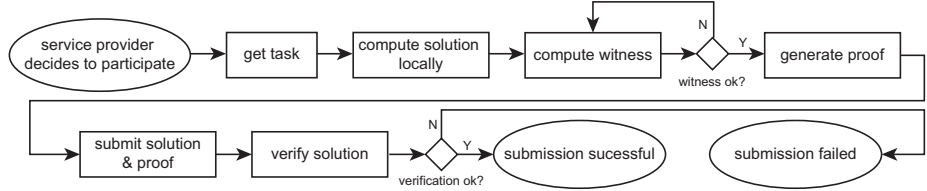}
	\caption{Result verification procedure}
	\label{fig:verificationProcess}
\end{figure}
The procedure of the result verification is illustrated in Figure~\ref{fig:verificationProcess}
. First, a service provider decides to contribute to a particular offloading task, \added{which is the start event for the result verification}
. For this, the provider retrieves all relevant input data. Based on that, the service provider computes the task locally (and therefore off-chain) with a service running on the provider's computational resources. After a solution has been found, a witness has to be computed. This means that a program specified in DSL code is run with the service provider's result as input; this DSL program is providing important input for the ZoKrates toolbox as discussed in more detail in Section~\ref{sub:example}. If the verification succeeds, the DSL program returns a witness that proves proper computation. Otherwise, it can be assumed that an error occurred during the computation phase or a wrong result has been entered. In this case, the computation is repeated. If the computation fails again, the procedure ends unsuccessfully (not shown in Fig.~\ref{fig:verificationProcess}).

Based on the witness, a proof can be generated, which is needed for the on-chain verification in the next step. Both actions, compute witness and generate proof are performed locally on the service provider's hardware. Afterwards, the solution and the proof can be submitted to the broker smart contract. When a solution is submitted, the broker smart contract calls the verification function \textit{verifyTx()} of the verification smart contract. If the verification function returns true, we can assume that the computation has been executed honestly. As a result, the solution is stored on-chain. Otherwise, if false is returned, the submission of the service provider is discarded entirely. \added{In both cases, the procedure subsequently ends.}

\subsection{\replaced{Implementing Result Verification for Specific Use Cases}{Implementation of the Result Verification}}
\label{subsec:implOfResultVerification}
Result verification is naturally tied to specific computation problems. This means that a certain part of the result verification, more specifically the DSL program implemented using the ZoKrates toolbox, is not generic and must be adapted whenever a different computation problem has to be served. 

\begin{figure}[t] 
	\centering
	\includegraphics[width=\textwidth]{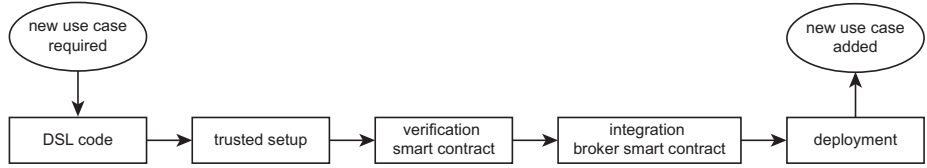}
	\caption{Process for adding result verification for additional use cases}
	\label{fig:addResultVerificationForNewUseCase}
\end{figure}

Accordingly, the DSL code contains appropriate checks to prove that a computation is done correctly. The possibility of creating or adopting these programs enables flexibility and adds universal applicability to the result verification. In other words, new use cases can be added and thus potential demands of service consumers for new computation problems can be met. The mandatory steps to add a new use case are depicted in Figure~\ref{fig:addResultVerificationForNewUseCase}.

To verify results of a specific computation problem, several one-time preparation steps have to be conducted. As indicated, a program in form of DSL code has to be written. \added{Within the DSL program, some logic, e.g., conditions, has to be specified, which makes the solution for a specific problem true. The DSL program takes a number of inputs (depending on the use case) and verifies if all specified conditions are met. Consider the offloading task of calculating the sum of two integers. In this case, the equation $a + b = c$ must hold and has to be encoded in the DSL. In general, an arbitrary task can be verified, as long as it is possible to define a rule that the result has to follow. As described in Section~\ref{sec:offchain}, for this, the service provider does not have to disclose any information about the applied algorithm or method.}

Then, keys and the verification smart contract have to be generated. For the keys, this means that a trusted setup is necessary. In the work at hand, we assume that the trusted setup is performed by the developers of the smart contracts presented in Section~\ref{sub:brokering}. Afterwards, the on-chain verification for the specific computation problem is ready for deployment. Alternatively, a multiparty computation protocol could be applied, e.g.,~\cite{DBLP:conf/uss/Ben-SassonCTV14}, that prevent fake proofs as long as one participant is honest.

When the result verification scheme is integrated into the broker smart contract, the inputs of the function~\textit{submitSolution()}, as well as the storage format for solutions of the new problem, should be kept in mind. Under certain circumstances, it can be useful to draw up a detached broker smart contract for each use case. In this case, the basic functions of the broker smart contract, e.g.,~\textit{getTaskRequest()} can be copied, but use case-specific data, e.g., necessary fields for the problem instance and solutions, must be considered and adapted as well. Furthermore, the function \textit{submitSolution()} that also calls \textit{verifyTx()} of the verification smart contract has to be tailored. 
A separation per use case would lead to more compact code artifacts and better maintenance. However, to simplify the description of our solution approach, we only discuss the usage of one broker smart contract in this paper.

\added{When a new use case is added, the main effort consists of rewriting the DSL program and integrating the verification smart contract into the broker smart contract. The aforementioned trusted setup and creation of the verification smart contract is mainly handled by the ZoKrates toolbox. The deployment of new smart contracts is tool-supported as well. This means that merely predefined commands have to be executed.}

\subsection{Example Implementation}
\label{sub:example}
To demonstrate how our blockchain-based offloading approach with result verification can be adopted to specific use cases, we describe the process of implementing an exemplary use case based on the TSP. 

The TSP is selected since it is very well-known among computer scientists, and easy to understand, but actually hard to compute. In brief, the TSP describes the problem to find the shortest path to travel a predefined number of cities and to return back to the origin city, but visiting any other city only once. As input, a list of cities and the distances between each pair of cities are given~\cite{applegate07}. Once a solution to the TSP has been found, its validity can be verified in little time, e.g., by traversing the path of a solution. \added{Interestingly, it is also easy to assess the quality of the solution, i.e., by comparing the computed path length.}

The TSP is an NP-hard problem, which also opens interesting future research direction (see Section~\ref{sec:conclusion}). Notably, the TSP is merely used as an exemplary use case in the work at hand. The presented approach explicitly allows to verify other tasks as well. 
The according implementation including the broker smart contract can be found at Github\footnote{\url{https://github.com/ben2048/blockchainBasedComputationOffloading}}.

There are some restrictions which have to be taken into account because of the applied ZoKrates toolbox. First, the toolbox only supports numbers, i.e., cities in the TSP have to be represented by numbers, not by their names. Second, ZoKrates does not support any dynamic fields. Accordingly, the size of an array needs to be defined at compile time. Both constraints complicate the implementation a little bit, but do not lead to any significant restrictions. We discuss the impact of the fixed array sizes also in the evaluation in Section~\ref{sec:evaluation}.

\renewcommand{\lstlistingname}{Alg.}
\begin{lstlisting}[
	xleftmargin=2em,
	xrightmargin=2em,
	frame=single,
	showspaces=false,
	basicstyle=\scriptsize\ttfamily,
	commentstyle=\color{Green}\ttfamily,
	numbers=left,
	numberstyle=\tiny\color{gray},
	caption=Main function of the DSL program,
	captionpos=t,
	label={lst:tspDslMain},
	tabsize=2,
	%aboveskip=1em,
	float
	]
	def main(
	private field[10] path, private field mapnumber, field sum, 
	private field[10] cities, field[2] hashOfCities, field[2] hashOfPath
	) -> (field):
	
	1 == basicInputCheck(path, cities, mapnumber)
	1 == checkCities(path, cities, mapnumber)
	sum == calculateSum(path, mapnumber)
	
	field[2] hashedPath = hash(concat(path))
	hashOfPath[0] == hashedPath[0]
	hashOfPath[1] == hashedPath[1]
	
	field[2] hashedCities = hash(concat(cities))
	hashOfCities[0] == hashedCities[0]
	hashOfCities[1] == hashedCities[1]
	
	return 1
\end{lstlisting}

As described in Section~\ref{subsec:implOfResultVerification}, the result verification is based on a DSL which proves program inputs against defined checks. \deleted{For example, if service providers intend to prove that they know the preimage of a hash, the DSL program has to check if the preimage hashes to the expected result
. Basically, arbitrary problems can be covered by such programs.} 
With regard to the TSP, the result verification has to verify whether a solution has been computed properly. In other words, service providers need to prove that they have found a valid solution for an TSP instance. To accomplish that, the produced path has to be Hamiltonian (i.e., each city appears exactly once in the path) and the path length must correspond to the sum of the connections between the cities on the basis of the path and the given distances~\cite{applegate07}. As input, the map of cities for the TSP has to be defined, including the cities (represented by numbers) and the distances between the cities, i.e., a complete graph made up from vertices (cities) and edges (distances) between the vertices. This data structure can be stored as an array within the DSL program.

In the following paragraphs, we discuss the example given in Algorithm~\ref{lst:tspDslMain}. To verify that a service provider has computed a correct result, the following information needs to be provided (lines 2--3): 
(i)~The computed \texttt{path}, consisting of a sequence of numbers representing cities, (ii)~an ID \texttt{mapnumber} for the map which has been used for solving the TSP instance (this allows to use different maps with the TSP), (iii)~the computed length of the path \texttt{sum}, (iv)~the \texttt{cities} for which the minimal distance has been computed, and (v)~the hashed path \texttt{hashOfPath} and the hashed cities \texttt{hashOfCities}. As it can be seen, the example allows a maximum of ten cities on a path, but other lengths are also possible. 

The path and the map are needed to calculate the distance and to compare it with the stated length (i.e., the result) from the service provider. Path and cities are used to determine if all cities are covered exactly once in a path. The hashed path is necessary to prevent malicious behavior originating from the service provider when submitting a solution. Without the hash, it would be possible to decouple the proof from the path. In other words, if the service provider submits a valid proof but an invalid path, it possibly cannot be detected on-chain, i.e., the result verification would succeed even though an incorrect path would be stored on the blockchain. To recognize and prevent such scenarios, we compare the hash and the path within the DSL program and in the broker smart contract.

As it can be seen, the main function first performs an input check (line 6). This is done in order to sort out solutions with invalid indices or map numbers. Next, it is checked if the stated path contains each city exactly once (line 7), i.e., if the path is Hamiltonian. Afterwards, it is checked if the stated path length (field \texttt{sum} in line~8) is equal to the sum resulting from the distances of the stated path based on the distances between cities, i.e., the map. Then, it is necessary to embed the hash of the path in the verification procedure (lines 10--12). Therefore, the hash of the path is needed as input. Consequently, we have to compute the hash of the stated path and compare it with the input, to prevent the aforementioned malicious action. For this, we utilize the implementation of SHA256 provided by ZoKrates. 

As discussed before, the DSL program shown in Alg.~\ref{lst:tspDslMain} runs off-chain, while the verification function \emph{verifyTx()} is carried out on-chain. Notably, the number of inputs of \emph{verifyTx()} is a cost factor. With regard to our use case and the specification of the input parameters in the DSL program, the number of cities scales with the instance size. This becomes a crucial cost factor, since the gas cost rise linearly with the number of public inputs provided. Hence, we make use of the hashed cities instead of the number of cities. This allows to make the cities a private input, but also makes it necessary to check the hash of the cities (lines 14--16), analogue to lines 10--12.

\begin{lstlisting}[
	xleftmargin=2em,
	xrightmargin=2em,
	frame=single,
	showspaces=false,
	basicstyle=\scriptsize\ttfamily,
	commentstyle=\color{Green}\ttfamily,
	numbers=left,
	numberstyle=\tiny\color{gray},
	caption=verifyTx(),
	captionpos=t,
	label={lst:verifytx},
	tabsize=2,
	%aboveskip=1em,
	float
	]
	function verifyTx(
	uint[2] memory a, uint[2][2] memory b, 
	uint[2] memory c, uint[6] memory input
	) public returns (bool r) {
		Proof memory proof;
		proof.a = Pairing.G1Point(a[0], a[1]);
		proof.b = Pairing.G2Point([b[0][0], b[0][1]], [b[1][0], b[1][1]]);
		proof.c = Pairing.G1Point(c[0], c[1]);
		uint[] memory inputValues = new uint[](input.length);
		for(uint i = 0; i < input.length; i++){
			inputValues[i] = input[i];
		}
		if (verify(inputValues, proof) == 0) {
			return true;
		} else {
			return false;
		}
	}
\end{lstlisting}

Based on the DSL program, ZoKrates is able to define \emph{verifyTx()} as depicted in Algorithm~\ref{lst:verifytx}. The input array consists of the path length, hash values for the path and the cities plus the expected return value of the DSL program (lines 2--3). Afterwards, a new proof is instantiated (line~5). Then, \emph{verifyTx()} requires three elliptic curve points in form of the arrays \texttt{a}, \texttt{b}, \texttt{c} (lines 6--8). These elliptic curve points actually make the zk-SNARKs proof and are delivered via the DSL program by the ZoKrates toolbox. Hence, the developer who integrates \emph{verifyTx()} does not have to take care of the actual proofs, which is a major reason for using the ZoKrates toolbox.

The array \texttt{input} depicts the public inputs and the expected return value of the DSL code, and is used to fill the \texttt{inputValues} (lines~9--12). Afterwards, the actual verification is carried out~(line 13). In the case of a successful verification, the boolean true is returned (line 14), else, the boolean false is returned (line 16)\deleted{, i.e., written to the blockchain}. Thus, a service consumer can be sure that a result is valid (or not), and the broker smart contract could carry out the payment. Notably, since only the proof is published on the blockchain, no conclusions regarding the computation and concrete results are possible. This ensures the privacy property regarding the proof. However, as written above, the solution is still available on the chain. To avoid this, a solution could be encrypted (see Section~\ref{sub:brokersmartcontract}).

\section{Evaluation}
\label{sec:evaluation}
\subsection{Evaluation Setup}
\label{sub:setup}
In order to evaluate the presented approach, we measure the overhead (regarding time and cost) occurring because of result verification. 

As mentioned in Section~\ref{sec:design}, the computation offloading consists of an on-chain and an off-chain part. To evaluate the on-chain activities, the smart contracts have been deployed in a local Ethereum blockchain (using Truffle), while for the off-chain activities, ZoKrates has been installed locally in a Docker container. 


\subsection{Overhead Analysis}
\label{sub:overhead}

The result verification is part of the solution submission process as presented in Sections~\ref{sub:brokersmartcontract} and \ref{sub:resultverification}. Hence, in order to evaluate the overhead with regard to gas cost and time, we implemented a second \emph{submitSolution()} function in the broker smart contract. However, this version of the function does \emph{not} verify the submitted solution. This allows us to compare the gas and time consumption of a benchmark with the according values of our zk-SNARKs-based verification approach. As a second benchmark, we conduct an on-chain result verification, i.e., a solution for a TSP instance is verified by a separate smart contract deployed on the blockchain. We apply two maps with size 30 and 70, respectively, in order to see how the map size (i.e., number of cities) influences the results. To get a complete picture of the cost overhead, TSP instance sizes between 3 and 30 (\textit{map 30}) respectively 3 and 60 (\textit{map 70}) are used.

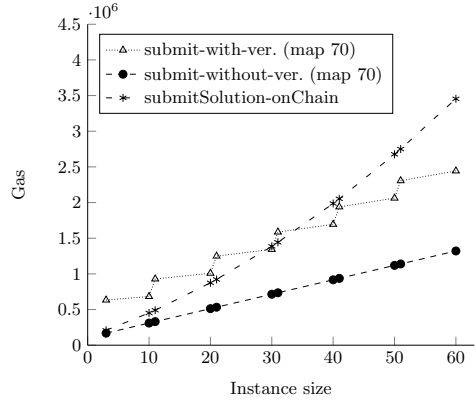
\begin{figure}[t]
	\centering
		\begin{tikzpicture}[scale=0.75]
			\begin{axis}[
				xlabel={Instance size},
				ylabel={Gas},
				xmin=0, xmax=63,
				ymin=0, ymax=4500000,
				axis x line*=bottom,
				axis y line*=left,
				xtick={0,10,20,30,40,50,60},
				ytick={0,500000,1000000,1500000,2000000,2500000,3000000, 3500000, 4000000, 4500000},
				legend pos=north west,
				legend style={legend cell align=left},
				]
				
				\addplot[color=black,mark=triangle,densely dotted, mark options={solid}]
				coordinates {
					(3,632 922)
					(10,683 095)
					(11,928 390)
					(20,1 008 588)
					(21,1 249 366)
					(30,1 343 625)
					(31,1 586 275)
					(40,1 694 698)
					(41,1 939 371)
					(50,2 061 946)
					(51,2 306 302)
					(60,2 442 993)
				};

				\addplot[color=black,mark=*,dashed, mark options={solid}]
				coordinates {
					(3,168 740)
					(10,310 252)
					(11,330 468)
					(20,512 413)
					(21,532 629)
					(30,714 575)
					(31,734 791)
					(40,916 736)
					(41,936 953)
					(50,1 118 899)
					(51,1 139 115)
					(60,1 321 061)};
				
				\addplot[color=black,mark=asterisk,loosely dashed, mark options={solid}]
				coordinates {
					(3,208 557)
					(10,451 771)
					(11,490 067)
					(20,874 698)
					(21,921 874)
					(30,1 386 423)
					(31,1 442 480)
					(40,1 986 947)
					(41,2 051 884)
					(50,2 676 271)
					(51,2 750 088)
					(60,3 454 393)
				};
				
				\legend{submit-with-ver. (map 70), submit-without-ver. (map 70), submitSolution-onChain}
			\end{axis}
		\end{tikzpicture}
		\captionof{figure}{Gas consumption}
		\label{fig:overheadSubmitSolution}
\end{figure}

The results regarding the cost for map 70 are shown in Figure~\ref{fig:overheadSubmitSolution}. Not surprisingly, the gas consumption with verification is higher than without verification. If no verification is done, there is no difference between the two maps. Hence, there is only one plot for the offloading without verification. Actually, this benchmark also indicates how big the cost become if redundant computation is used (see Section~\ref{sec:related}). In that case, based on the level of redundancy, each submitted solution leads to the same cost. In addition, some overhead for the comparison (e.g., majority voting) of the results needs to be taken into account. This shows that redundant computing is not really an option with regard to the gas consumption for the solution submission.

While not shown in the figure in order not to overload the plot, the submission of solutions based on the smaller map (with 30 cities) is marginally cheaper than solutions based on the larger map (with 70 cities). 

It can also be seen that verifying TSP solutions on-chain (\emph{submitSolution-onChain}) is cheaper than the approach presented in this work for small instances, but more expensive for large ones. Up to an instance size of 29, the on-chain verification is cheaper than applying zk-SNARKs. For instances of size 30, the \replaced{zk-SNARKs-based}{constant} version should be preferred, whereby on-chain verification is at a lower price for an instance size of 31. Finally, the cost of on-chain verification exceeds the cost of the zk-SNARKs-based variant at instance size 40. Overall, we can clearly see which version is cheaper for instances of size up to 29 and from 40 ascending. Around the instance size 30, three intersections with regard to the on-chain and zk-SNARKs-based variant are visible. To determine the exact break-even point(s) between 29 and 40, we have performed further (not-depicted) measurements showing that the on-chain variant is more cost-efficient for instances of size 31--34. From instance size 35, the zk-SNARKs-based variant should be preferred.

These results show that it is necessary to discuss the line course in more detail: It becomes clear that the gas demand of all three variants shown in Figure~\ref{fig:overheadSubmitSolution} increases with the instance size. 
It is also noticeable that the cost levels of \textit{submit-without-ver. (map 70)} and \emph{submitSolution-onChain} rise continuously, while the levels of \textit{submit-with-ver. (map 70)} increase step-wise. This is caused by the partitioned verification, due to the lack of dynamic fields and the subsequent fixed array sizes within the DSL of ZoKrates. For solutions of size 3 to 60, separate DSL programs with a varying number of inputs (starting with an instance size of 10, and increased by steps of 10) are provided. For example, if a DSL solution for 11 cities is computed, the resulting path has to be padded, according to the expected number of inputs of the DSL program. \deleted{Therefore, in our example, the path within the solution needs nine placeholders and has consequently a length of 20.} Due to the fact that such a padding is not necessary when no verification is performed or the verification is done on-chain, the gas consumption merely rises continuously for these options, but for the zk-SNARKs-based approach, ``jumps'' in the plot can be seen. A solution to circumvent this would be to have different smart contracts for different TSP instance sizes. \deleted{In fact, it can be assumed that in real-world deployments, separate result verifications for different problems would have to be implemented (see Section~\ref{sec:design}).}

\begin{figure}[t]
	\centering
	\begin{tikzpicture}[scale=0.75]
		\begin{axis}[
			ybar stacked,
			width=14cm, height=7cm,
			legend pos=north west,
			axis x line*=bottom,
			axis y line*=left,
			legend style={legend cell align=left},
			ylabel={running time [seconds]},
			xlabel={Instance size/map},
			x label style={below=7mm},
			ytick={0,25,50,75,100,125,150,175,200},
			symbolic x coords={3/m30, 3/m70, 10/m30, 10/m70, 11/m30, 11/m70, 20/m30, 20/m70, 21/m30, 21/m70, 30/m30, 30/m70, 31/m70, 40/m70, 41/m70, 50/m70, 51/m70, 60/m70},
			xtick=data,
			x tick label style={rotate=45,anchor=east},
			]
			\addplot+[ybar,black,fill=lightgray] plot coordinates {
				(3/m30,5.783)
				(10/m30,5.835)
				(11/m30,11.075)
				(20/m30,10.795)
				(21/m30,17.06)
				(30/m30,16.752)
				
				(3/m70,5.907)
				(10/m70,5.707)
				(11/m70,10.767)
				(20/m70,12.348)
				(21/m70,17.611)
				(30/m70,16.77)
				(31/m70,26.531)
				(40/m70,28.591)
				(41/m70,34.472)
				(50/m70,34.609)
				(51/m70,39.545)
				(60/m70,49.516)
			};
			\addplot+[ybar,black,fill=white,postaction={pattern=north east lines,pattern color=gray}] plot coordinates {
				(3/m30,18.428)
				(10/m30,18.305)
				(11/m30,36.111)
				(20/m30,36.242)
				(21/m30,54.026)
				(30/m30,54.681)
				
				(3/m70,18.673)
				(10/m70,18.08)
				(11/m70,37.304)
				(20/m70,41.373)
				(21/m70,57.368)
				(30/m70,53.739)
				(31/m70,82.885)
				(40/m70,89.698)
				(41/m70,103.956)
				(50/m70,107.994)
				(51/m70,170.602)
				(60/m70,152.356)
			};
			
			\legend{compute-witness, generate-proof}
		\end{axis}
	\end{tikzpicture}
	\caption{Time overhead} \label{fig:offchainPerformance}	
\end{figure}
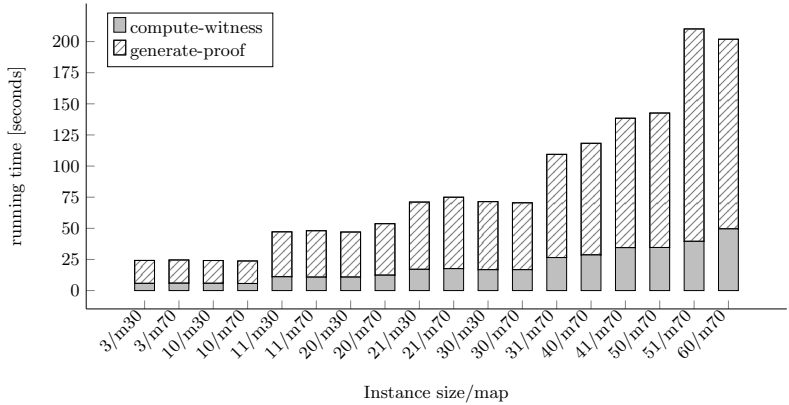

Second, we observe the time overhead. Figure~\ref{fig:offchainPerformance} depicts the time needed to execute the \emph{compute-witness} and \emph{generate-proof} step for TSP instances of sizes 3 to 60. The values on the x-axis can be interpreted as follows: 3/m30 means that the instance is of size three and belongs to map 30. As can be seen, the overall run-time increases with the size of instances. The step \emph{generate-proof} takes on average 3.2 (standard dev.: 0.28) times longer than the \emph{compute-witness} step. Considering the increasing runtime of the proof generation and the computation of the witness, it becomes clear that the presented result verification approach should primarily be used in scenarios which are not very time-critical.



\section{Conclusion}
\label{sec:conclusion}
In order to provide solutions for fully decentralized task offloading, a number of blockchain-based solutions have already been proposed. However, to the best of our knowledge, none of these solutions is able to verify that an offloaded task is computed correctly. Instead, redundant computing or trust models are applied, which however cannot guarantee that a computation is valid. 

We have therefore presented an approach which supports the verification of computation results, applying zk-SNARKs. Especially, this allows to compute results and proofs off-chain, and to only verify the proofs on-chain. 
While our solution can also be integrated into existing blockchain-based offloading frameworks, we have also provided a simplified broker solution as part of this paper.

In our future work, we want to further extend the presented approach. Especially, we want to replace the brokering functionality by a more sophisticated one which also allows the broker to take into account quality requirements (e.g., a particular quality of a result), and to select based on this the best result from a number of provided solutions. In fact, selecting the TSP as our evaluation use case already lays the foundations for this, since service providers could deliver different solution qualities with different algorithms and at different cost to this NP-hard problem. In other scenarios, the assessment of the result quality is a more complex task and therefore an interesting direction of future work. 

While currently our reference implementation applies a simple pricing scheme, i.e., the service provider is paid with the consumer's stake, more complex pricing might be useful. For instance, a dynamic pricing scheme based on the complexity of an offloaded computational task and the number of available service providers might be helpful.
Last but not least, as has been shown in the evaluation, an on-chain verification is sometimes cheaper to conduct than the proposed off-chain result verification. Therefore, we will further investigate in which cases which of these two approaches should be preferred. 

\subsubsection*{Acknowledgments.} The financial support by the Austrian Federal Ministry for Digital and Economic Affairs, the National Foundation for Research, Technology and Development and the Christian Doppler Research Association is gratefully acknowledged.
%
%
%
%
\bibliographystyle{splncs04}
\bibliography{references}

\begin{thebibliography}{10}
\providecommand{\url}[1]{\texttt{#1}}
\providecommand{\urlprefix}{URL }
\providecommand{\doi}[1]{https://doi.org/#1}

\bibitem{applegate07}
Applegate, D.L., Bixby, R.E., Ch\'{a}tal, V., Cook, W.J.: {The Traveling
  Salesman Problem -- A Computational Study}. Princeton University Press (2007)

\bibitem{DBLP:conf/uss/Ben-SassonCTV14}
Ben{-}Sasson, E., Chiesa, A., Tromer, E., Virza, M.: {Succinct Non-Interactive
  Zero Knowledge for a von Neumann Architecture}. In: 23rd {USENIX} Security
  Symp. pp. 781--796. USENIX Association (2014)

\bibitem{DBLP:journals/tmc/ChatzopoulosAKH18}
Chatzopoulos, D., Ahmadi, M., Kosta, S., Hui, P.: {FlopCoin: {A} Cryptocurrency
  for Computation Offloading}. {IEEE} T. on Mobile Computing  \textbf{17}(5),
  1062--1075 (2018)

\bibitem{DBLP:conf/middleware/EberhardtH18}
Eberhardt, J., Heiss, J.: {Off-chaining Models and Approaches to Off-chain
  Computations}. In: 2nd Works. on Scalable and Resilient Infrastructures for
  Distributed Ledgers. pp. 7--12. ACM (2018)

\bibitem{DBLP:conf/ithings/EberhardtT18}
Eberhardt, J., Tai, S.: {ZoKrates -- Scalable Privacy-Preserving Off-Chain
  Computations}. In: 1st {IEEE} Int. Conf. on Blockchain. pp. 1084--1091. IEEE
  (2018)

\bibitem{iexecWhitepaper}
Fedak, G., Bendella, W., Alves, E.: {Blockchain-Based Decentralized Cloud
  Computing}.
  \url{https://iex.ec/wp-content/uploads/pdf/iExec-WPv3.0-English.pdf} (2017)

\bibitem{DBLP:conf/stoc/GoldreichMW87}
Goldreich, O., Micali, S., Wigderson, A.: {How to Play any Mental Game or {A}
  Completeness Theorem for Protocols with Honest Majority}. In: 19th Annual
  {ACM} Symp. on Theory of Computing. pp. 218--229. ACM (1987)

\bibitem{intel2019Plundervolt}
Greenberg, A.: {Hackers Can Mess With Voltages to Steal Intel Chips' Secrets}.
  \url{https://www.wired.com/story/plundervolt-intel-chips-sgx-hack}

\bibitem{kosba20}
Kosba, A., Papadopoulos, D., Papamanthou, C., Song, D.: {MIRAGE: Succinct
  Arguments for Randomized Algorithms with Applications to Universal
  zk-SNARKs}. In: 29th {USENIX} Security Symp. pp. 2129--2146. {USENIX}
  Association (2020)

\bibitem{kosta12}
Kosta, S., Aucinas, A., Hui, P., Mortier, R., Zhang, X.: {ThinkAir: Dynamic
  resource allocation and parallel execution in the cloud for mobile code
  offloading}. In: 31st IEEE Int. Conf. on Computer Communications. pp.
  945--953. {IEEE} (2012)

\bibitem{KSS20}
Krejci, S., Sigwart, M., Schulte, S.: {Blockchain- and IPFS-based Data
  Distribution for the Internet of Things}. In: 8th Europ. Conf. on
  Service-Oriented and Cloud Computing. LNCS, vol. 12054, pp. 177--191.
  Springer (2020)

\bibitem{mach17}
Mach, P., Becvar, Z.: {Mobile Edge Computing: {A} Survey on Architecture and
  Computation Offloading}. {IEEE} Communications Surveys \& Tutorials
  \textbf{19}(3),  1628--1656 (2017)

\bibitem{DBLP:conf/hicss/MarstonLBG11}
Marston, S., Li, Z., Bandyopadhyay, S., Zhang, J., Ghalsasi, A.: {Cloud
  Computing -- The Business Perspective}. In: Decision Support Systems.
  vol.~51, pp. 176--189 (2011)

\bibitem{DBLP:journals/iotj/PanWHALZ19}
Pan, J., Wang, J., Hester, A., AlQerm, I., Liu, Y., Zhao, Y.: {EdgeChain: An
  Edge-IoT Framework and Prototype Based on Blockchain and Smart Contracts}.
  {IEEE} Internet of Things J.  \textbf{6}(3),  4719--4732 (2019)

\bibitem{DBLP:journals/tvt/QiuLCHZ19}
Qiu, X., Liu, L., Chen, W., Hong, Z., Zheng, Z.: {Online Deep Reinforcement
  Learning for Computation Offloading in Blockchain-Empowered Mobile Edge
  Computing}. {IEEE} T. on Vehicular Technology  \textbf{68}(8),  8050--8062
  (2019)

\bibitem{khan14}
ur~Rehman~Khan, A., Othman, M., Madani, S.A., Khan, S.U.: {A Survey of Mobile
  Cloud Computing Application Models}. {IEEE} Communications Surveys \&
  Tutorials  \textbf{16}(1),  393--413 (2014)

\bibitem{santos09}
Santos, N., Gummadi, K.P., Rodrigues, R.: {Towards trusted cloud computing}.
  In: 2009 Conf. on Hot Topics in Computing. {USENIX} Association (2009),
  article No. 3

\bibitem{DBLP:journals/sigmobile/Satyanarayanan14}
Satyanarayanan, M.: {A Brief History of Cloud Offload: {A} Personal Journey
  from Odyssey Through Cyber Foraging to Cloudlets}. GetMobile: Mobile
  Computing and Communications  \textbf{18}(4),  19--23 (2014)

\bibitem{DBLP:journals/corr/SchwarzWGMM17}
Schwarz, M., Weiser, S., Gruss, D., Maurice, C., Mangard, S.: {Malware Guard
  Extension: Using {SGX} to Conceal Cache Attacks}. In: Int. Conf. on Detection
  of Intrusions and Malware, and Vulnerability Assessment. LNCS, vol. 10327,
  pp. 3--24. Springer (2017)

\bibitem{golemArchitecture}
Skrzypczak, A.: {Golem Architecture}.
  \url{https://blog.golemproject.net/golem-architecture/} (2017), online, last
  visited at 2021-05-20

\bibitem{sonmWhitepaper}
{Sonm Pte. Ltd.}: {SONM -- Supercomputer Organized by Network Mining}.
  \url{https://whitepaper.io/document/326/sonm-whitepaper}

\bibitem{DBLP:journals/corr/abs-1908-04756}
Teutsch, J., Reitwie{\ss}ner, C.: A scalable verification solution for
  blockchains. CoRR  \textbf{abs/1908.04756} (2019),
  \url{http://arxiv.org/abs/1908.04756}

\bibitem{DBLP:journals/csm/UriarteN18}
Uriarte, R.B., {De Nicola}, R.: {Blockchain-Based Decentralized Cloud/Fog
  Solutions: Challenges, Opportunities, and Standards}. {IEEE} Communications
  Standards Magazine  \textbf{2}(3),  22--28 (2018)

\bibitem{wu20}
Wu, H., Sun, Y., Wolter, K.: {Energy-Efficient Decision Making for Mobile Cloud
  Offloading}. {IEEE} T. on Cloud Computing  \textbf{8}(2),  570--584 (2020)

\bibitem{wu21}
Wu, H., Wolter, K., Jiao, P., Deng, Y., Zhao, Y., Xu, M.: {{EEDTO:} An
  Energy-Efficient Dynamic Task Offloading Algorithm for Blockchain-Enabled
  IoT-Edge-Cloud Orchestrated Computing}. {IEEE} Internet of Things J.
  \textbf{8}(4),  2163--2176 (2021)

\end{thebibliography}
%
\end{document}